\documentclass[pre,amsmath,amssymb,groupedaddress,graphicx,showkeys,showpacs,floatfix,reprint]{revtex4-1}

\usepackage[colorlinks=true,linkcolor=blue,citecolor=blue,urlcolor=blue,bookmarks,breaklinks=true]{hyperref}
\usepackage{breakurl}
\usepackage{url}
\usepackage{units}
\usepackage{graphicx}
\usepackage{subfigure}
\usepackage[utf8]{inputenc}
\usepackage{ae}
\usepackage{natbib}
\usepackage{dcolumn}
\usepackage{multirow}

\renewcommand{\vec}[1]{\boldsymbol{#1}}
\newcolumntype{d}{D{.}{.}{7}}

\begin{document}

\title{Characterization of maximally random jammed sphere packings:\\
  Voronoi correlation functions}

\author{Michael A. Klatt}
\affiliation{Department of Chemistry, Department of Physics, Princeton
  University, Princeton, New Jersey 08544, USA}
\affiliation{Friedrich-Alexander-Universit\"at Erlangen-N\"urnberg (FAU),
  Institut f\"ur Theoretische Physik, Staudtstraße 7, 91058 Erlangen,
  Germany}
\author{Salvatore Torquato}
\email[Electronic mail: ]{torquato@electron.princeton.edu}
\affiliation{Department of Chemistry, Department of Physics, Princeton
  Institute for the Science and Technology of Materials, and Program
  in Applied and Computational Mathematics, Princeton University, Princeton, New Jersey 08544, USA}
\date{\today}

\begin{abstract}
  We characterize the structure of maximally random jammed (MRJ)
  sphere packings by computing the Minkowski functionals (volume,
  surface area, and integrated mean curvature) of their associated
  Voronoi cells. The probability distribution functions of these
  functionals of Voronoi cells in MRJ sphere packings are
  qualitatively similar to those of an equilibrium hard-sphere liquid
  and partly even to the uncorrelated Poisson point process, implying
  that such local statistics are relatively structurally
  insensitive. This is not surprising because the Minkowski
  functionals of a single Voronoi cell incorporate only local
  information and are insensitive to global structural information. To
  improve upon this, we introduce descriptors that incorporate
  nonlocal information via the correlation functions of the Minkowski
  functionals of two cells at a given distance as well as certain
  cell-cell probability density functions. We evaluate these
  higher-order functions for our MRJ packings as well as equilibrium
  hard spheres and the Poisson point process. It is shown that these
  Minkowski correlation and density functions contain visibly more
  information than the corresponding standard pair-correlation
  functions. We find strong anticorrelations in the Voronoi volumes
  for the hyperuniform MRJ packings, consistent with previous findings
  for other pair correlations [A. Donev \textit{et al.},
    Phys. Rev. Lett. 95, 090604 (2005)], indicating that large-scale
  volume fluctuations are suppressed by accompanying large Voronoi
  cells with small cells, and vice versa. In contrast to the
  aforementioned local Voronoi statistics, the correlation functions
  of the Voronoi cells qualitatively distinguish the structure of MRJ
  sphere packings (prototypical glasses) from that of not only the
  Poisson point process but also the correlated equilibrium
  hard-sphere liquids. Moreover, while we did not find any perfect
  icosahedra (the locally densest possible structure in which a
  central sphere contacts 12 neighbors) in the MRJ packings, a
  preliminary Voronoi topology analysis indicates the presence of
  strongly distorted icosahedra.
\end{abstract}

\pacs{}{}
\maketitle

\section{Introduction}
\label{Introduction}

Packings of frictionless monodisperse hard spheres in three dimensions
serve as a simple, yet effective tool for modeling the complex
behavior of such diverse many-particle systems as crystals, colloids,
liquids, glasses, heterogeneous materials, foams, and biological
systems~\cite{ChaikinLubensky2000, ManoharanEtAl2003,
  HansenMcDonald2006, Torquato2002, KraynikEtAl2003, Zohdi2006,
  MejdoubiBrosseau2007, Sahimi2011, GevertzTorquato2008,
  GillmanEtAl2013}. Among the rich multitude of states they are known
to exhibit, considerable interest has been given towards sphere
packings that are jammed (i.e.,~mechanically stable), including
maximally dense packings, low-density crystals, and amorphous
packings~\cite{Henley1986, OHernLiuNagel2004, Hales2005, AsteEtAl2005,
  AsteMatteo2008, TorquatoStillinger2007, Finney1997, Zallen1998,
  TorquatoStillinger2010RevModPhys, KapferEtAl2012}.

In order to characterize the properties of sphere packings, one may
employ a geometric-structure approach in which configurations are
considered independently of both their frequency of occurrence and the
algorithm by which they are
created~\cite{TorquatoStillinger2010RevModPhys}.
For example, the simplest characteristic of a sphere packing is its
packing fraction $\phi$, i.e., the fraction of space occupied by the
spheres. Other useful characteristics of the structure include its
pair-correlation function~\cite{ClarkeJonsson1993, SeidlerEtAl2000,
  WeeksWeitz2002, AsteEtAl2005, TorquatoStillinger2006,
  TorquatoJiao2010, DelaneyEtAl2010, JiaoStillingerTorquato2011,
  PalomboGabrielliEtAl2013}, the pore-size
distribution~\cite{Torquato2002, DonevEtAl2005}, and structure
factor~\cite{HansenMcDonald2006, TorquatoStillinger2006,
  SilbertSilbert2009, KuritaWeeks2010, BerthierEtAl2011}.
 
It is also valuable to quantify the degree of ordering in a packing,
especially those that are jammed (defined more precisely below). To
this end, a variety of scalar order metrics $\psi$ have been
suggested~\cite{TorquatoEtAl2000PhysRevLetRCPvsMRJ,
  TorquatoStillinger2010RevModPhys} in which $\psi=0$ is defined as
the most disordered state (i.e.\ the Poisson point process) and
$\psi=1$ is the most ordered state. Using the geometric-structure
approach, one may therefore construct an ``order map'' in the
$\phi$--$\psi$ plane~\cite{TorquatoStillinger2010RevModPhys}, where
the jammed packings form a subset in this map. The boundaries of the
jammed region are optimal in some sense, including, for example, the
densest packings (the face-centered-cubic crystal and its stacking
variants with $\phi_{\mathrm{max}}=\pi/\sqrt{18}\approx
0.74$~\cite{Hales2005}) and the least dense jammed packings
(conjectured to be the ``tunneled crystals'' with
$\phi_{\mathrm{min}}=2\phi_{\mathrm{max}}/3$~\cite{TorquatoStillinger2007}).

Among the set of all isotropic and statistically homogeneous jammed
sphere packings, the maximally random jammed (MRJ) state is that which
minimizes some order metric $\psi$. This definition makes
mathematically precise the familiar notion of random closed packing
(RCP) in that it can be unambiguously identified for a particular
choice of order metric. A variety of sensible, positively correlated
order metrics produce an MRJ state with the same packing fraction
0.64~\cite{KansalEtAl2002,bibfootnote1}, which agrees roughly with the
commonly suggested packing fraction of RCP in three
dimensions~\cite{TorquatoStillinger2010RevModPhys}. However, we stress
that density alone is not sufficient to characterize a packing; in
fact, packings with a large range of $\psi$ may be observed at this
packing fraction~\cite{KansalEtAl2002,JiaoStillingerTorquato2011}.

In order to study the MRJ state, a precise definition of jamming is
needed. Therefore, rigorous hierarchical jamming categories have been
defined~\cite{TorquatoStillinger2001, TorquatoDonevStillinger2003}: A
packing is \textit{locally jammed} if no particle can move while the
positions of the other particles are fixed; it is \textit{collectively
  jammed} if no subset of particles can move without deforming the
system boundary; and if also a deformation of the system boundary is
not possible without increasing its volume, the packing is
\textit{strictly jammed}, i.e., it is stable against both uniform
compression and shear deformations~\footnote{The jamming category of a
  finite system will depend on the boundary conditions; see Refs.
  \cite{TorquatoStillinger2001} and
  \cite{TorquatoDonevStillinger2003}}. Strictly jammed MRJ sphere
packings often contain a small fraction of rattlers (unjammed
particles), but the remainder of the packing, i.e., the mechanically
rigid backbone, is strictly jammed~\cite{AtkinsonEtAl2013}.

Determining the contact network of a packing is a subtle problem
requiring high numerical fidelity.
The Torquato-Jiao (TJ) sphere packing
algorithm~\cite{TorquatoJiao2010} meets this challenge by efficiently
producing highly disordered, strictly jammed packings with unsurpassed
numerical fidelity as well as ordered packings~\cite{TorquatoJiao2010,
  MarcotteTorquato2013}. The algorithm achieves this by solving a
sequence of linear programs which iteratively densify a collection of
spheres in a deformable periodic cell subject to locally linearized
nonoverlap constraints. The resulting packings are, by definition,
strictly jammed and they are with high probability exactly isostatic
(meaning that they possess the minimum number of contacts required for
jamming)~\cite{DonevEtAl2007, AtkinsonEtAl2013}. The TJ packing
protocol is intrinsically capable of producing MRJ states with very
high fidelity~\cite{TorquatoJiao2010, AtkinsonEtAl2013}.
The MRJ state can be regarded as a prototypical glass because it is
maximally disordered (according to a variety of order metrics) while
having infinite elastic
moduli~\cite{TorquatoStillinger2010RevModPhys}.
Atkinson \textit{et al.}~\cite{AtkinsonEtAl2013} recently
carried out a detailed characterization of the rattler population in
these MRJ sphere packings. They found a rattler fraction of 1.5\,\%
(substantially lower than other packing protocols, such as the
well-known Lubachevsky-Stillinger packing
algorithm~\cite{LubachevskyStillinger1990}) and that the rattlers were
highly spatially correlated, implying that they have a significant
influence on the structure of the packing~\cite{DonevEtAl2005}.
Moreover, as in previous studies~\cite{DonevEtAl2007}, it was
shown~\cite{AtkinsonEtAl2013} that the backbone of the MRJ state is
isostatic. We include rattlers in our analysis unless stated
otherwise.

\begin{figure}[t]
  \centering
  \includegraphics[width=0.975\linewidth]{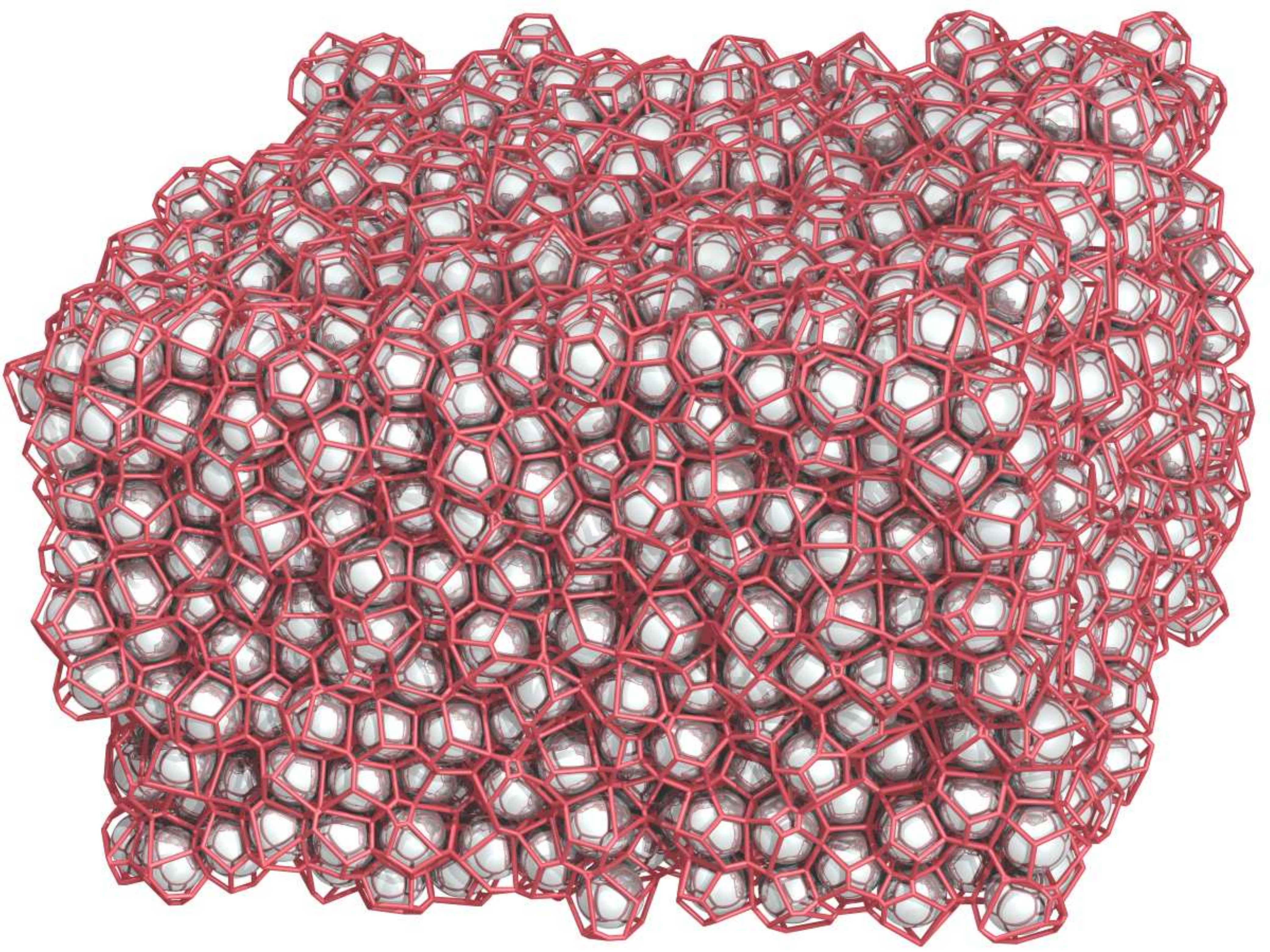}
  \caption{(Color online) Maximally random jammed (MRJ) sphere packing
    and its Voronoi diagram. Among all jammed sphere packings (roughly
    speaking, the mechanically stable packings), the MRJ state is the
    most disordered one.}
  \label{fig:exemplary_MRJ}
\end{figure}

MRJ packings have been characterized using a variety of statistical
descriptors, including the radial pair-correlation function $g_2(r)$
($\rho^2g_2$ is the probability density for finding two sphere centers
separated by a radial distance $r$, where $\rho$ is the number
density, i.e., the number of particles per unit
volume)~\cite{TorquatoJiao2010}, the bond-orientational order metric
$Q_6$ and the translational order metric $T^*$~\cite{KansalEtAl2002,
  AtkinsonEtAl2013}, the cumulative pore-size distribution
$F(\delta)$~\cite{DonevEtAl2005}, and the statistics of
rattlers~\cite{AtkinsonEtAl2013}.
In addition, MRJ sphere packings exhibit disordered
hyperuniformity~\cite{DonevEtAl2005}, meaning that they are locally
disordered, but possess a hidden order on large length scales such
that infinite-wavelength density fluctuations of MRJ packings vanish,
i.e., the structure factor vanishes at the origin:
$\lim_{\boldsymbol{k} \rightarrow \boldsymbol{0}} S(\boldsymbol{k}) =
0$~\cite{TorquatoStillinger2003,bibfootnote0}. Disordered
hyperuniformity can be seen as an ``inverted critical phenomenon'' with
a direct correlation function $c(r)$ that is
long ranged~\cite{TorquatoStillinger2003,
  HopkinsStillingerTorquato2012}.

In this paper, we characterize the MRJ sphere packings generated in
Ref.~\cite{AtkinsonEtAl2013} using Voronoi statistics, including
certain types of correlation functions. We compare these computations
to corresponding calculations for both a Poisson distribution
of points and equilibrium hard-sphere liquids. In the second paper of this
series, we will investigate density fluctuations, the pore-size
distribution, and two-point probability functions of MRJ packings.

Many studies for disordered sphere packings have been devoted to
computing the volume distribution of the Voronoi
cells~\cite[e.g.,][]{Finney1970, Finney1997, StarrEtAl2002,
  LechenaultEtAl2006, AsteMatteo2007, AsteMatteo2008,
  SchroederTurkEtAl2010, KapferEtAl2010, ZhaoEtAl2012, KapferEtAl2012}; see
Fig.~\ref{fig:exemplary_MRJ} for a MRJ sphere packing and its Voronoi
diagram~\footnote{For each sphere, a Voronoi cell is assigned which
  contains all points closer to this sphere than to any other sphere
  in the packing.}.
However, such statistics are incomplete in that they only quantify
local structural information.
For example, with appropriately rescaled variables, we will show that
the distributions of the Voronoi volumes, surface areas, and
integrated mean curvatures for the MRJ sphere packings are
qualitatively similar to the distributions for an equilibrium
hard-sphere liquid and partly even for the spatially uncorrelated
Poisson point process.

\begin{table*}[t]%
  \caption{\label{tab:minkfunc}%
    Mean $\langle W_{\mu}\rangle$, standard deviation
    $\sigma_{W_{\mu}}$, and correlation coefficients $\rho_{\mu,\nu}$ of the Minkowski
    functionals $W_{\mu}$ of single Voronoi cells in the Poisson
    point process, in a system of hard spheres in equilibrium at a
    packing fraction $\phi=0.48$, and in the MRJ state. The
    unit of length is $\lambda=1/\rho^{1/3}$, i.e., the number density
    $\rho$ is set to unity.}
  \centering
  \begin{ruledtabular}
    \begin{tabular}{l c d d d d}
      &  & \multicolumn{1}{c}{$\langle W_{\mu} \rangle$} & \multicolumn{1}{c}{$\sigma_{W_{\mu}}$} & \multicolumn{1}{c}{$\rho_{\mu,1}$} & \multicolumn{1}{c}{$\rho_{\mu,2}$} \\
      \hline
      &  & \multicolumn{4}{c}{Poisson} \\
      Volume          & $W_0$ & 1.0005(3) & 0.4230(2) & 0.98161(3) & 0.94486(8)\\
      Surface area    & $W_1$ & 5.823(1)  & 1.4798(7) & & 0.98701(2) \\
      Integ. mean curv.   & $W_2$ & 9.1623(8) & 1.0941(5) & & \\
      &  & \multicolumn{4}{c}{Equilibrium} \\
      Volume          & $W_0$ & 1.00000(7) & 0.07434(5) & 0.97700(5) & 0.9282(2)\\
      Surface area    & $W_1$ & 5.4488(3)  & 0.2681(2)   & & 0.98136(5)\\
      Integ. mean curv.   & $W_2$ & 8.6477(2)  & 0.2163(2)   & & \\
      &  & \multicolumn{4}{c}{MRJ} \\
      Volume          & $W_0$ & 1.00000(3) & 0.04335(2) & 0.96976(4) & 0.9035(1)\\
      Surface area    & $W_1$ & 5.4043(1)  & 0.1695(8)  & & 0.97248(5)\\
      Integ. mean curv.   & $W_2$ & 8.5894(1)  & 0.1470(7)  & &
    \end{tabular}
  \end{ruledtabular}
\end{table*}

To quantify nonlocal structural information, we formulate and compute
correlation functions of the volume of Voronoi cells at a given
distance and cell-cell probability density functions of finding a given
sized Voronoi cell at a given distance of a sphere with another sized
Voronoi cell.
Because the volume is only one of a large class of versatile shape
measures, namely, the Minkowski
functionals~\cite{SchneiderWeil2008,Mecke1998,MeckeStoyan2000,
  SchroederTurketal2010AdvMater}, we devise and compute the
correlation functions of all of the Minkowski
functionals~\footnote{The Euler characteristic, another Minkowski
  functional, is a topological constant. For a single Voronoi cell, it
  is always trivially equal to one.}. Besides characterizing MRJ packings
in this way, we also carry out analogous calculations for the  Poisson
point process and the equilibrium hard-sphere liquid for purposes of comparison.
We show that these Minkowski
correlation functions contain visibly more information than
the corresponding standard pair-correlation functions, even in the case
of the Poisson point process.

In Sec.~\ref{sec:MF_distributions}, we analyze the distributions of
the Minkowski functionals of the single Voronoi cells for the  Poisson
point process, equilibrium hard-sphere liquid configurations, and MRJ packings.
In Sec.~\ref{sec:definitions}, we define the aforementioned two
different types of correlation functions.
In Sec.~\ref{sec:3d}, we determine the volume-volume correlation
function numerically for the three-dimensional Poisson point process,
equilibrium hard-sphere systems, and the MRJ state; we also
calculate the correlation functions for the surface area, and the
integrated mean curvature.
For a further investigation of the nonlocal structure features, we
calculate in Sec.~\ref{sec:radial_prob} the cell-cell probability
density functions mentioned above. In Sec.~\ref{sec:Conclusion}, we
make concluding remarks.

\section{Minkowski Functional Distributions of a Single Voronoi Cell}
\label{sec:MF_distributions}

While there are many detailed studies of the volume distribution in
disordered sphere packings~\cite[e.g.,][]{Finney1970, Finney1997,
  StarrEtAl2002, LechenaultEtAl2006, AsteMatteo2007, AsteMatteo2008,
  SchroederTurkEtAl2010, KapferEtAl2010, ZhaoEtAl2012,
  KapferEtAl2012}, here we
analyze in a logarithmic plot the volume distributions of true MRJ packings as
describe above and extend the analysis to all three (nontrivial)
Minkowski functionals: the volume, the surface area, and the
integrated mean curvature.
They are robust and versatile shape descriptors which are widely used
in statistical physics~\cite{Mecke1998, MeckeStoyan2000,
  SchroederTurketal2010JoM, SchroederTurketal2010AdvMater} and in
pattern analysis~\cite{Mecke1996, Becker2003, MantzJacobsMecke2008}.
We use \textsc{voro++}~\cite{Rycroft2006, Rycroft2009} to construct
the Voronoi diagram of Poisson point patterns (about 1000 patterns,
each with 2000 points), equilibrium hard-sphere
packings~\cite{Torquato2002, HansenMcDonald2006} at a packing fraction
$\phi=0.48$, which is slightly below the freezing
transition (100 packings, each with 10000 spheres), and MRJ sphere
packings produced by the TJ algorithm~\cite{TorquatoJiao2010,
  AtkinsonEtAl2013} (about 1000 packings, each with 2000 spheres).
The program \textsc{karambola}~\cite{Mickel2008,
  SchroederTurkEtAl2013} then computes the Minkowski functionals of
each cell.

We first determine the distributions of the three Minkowski
functionals ($W_0$, volume; $W_1$, surface area; and $W_2$, integrated
mean curvature) of the single Voronoi cells. Table~\ref{tab:minkfunc}
provides the mean, the standard deviation, and the correlation
coefficients $\rho_{\mu,\nu}=\frac{\langle
  W_{\mu}W_{\nu}\rangle-\langle W_{\mu} \rangle\langle
  W_{\nu}\rangle}{\sigma_{W_{\mu}}\sigma_{W_{\nu}}}$ of the three
different Minkowski functionals.
As a unit of length, we use $\lambda=1/\rho^{1/3}$ with $\rho$ the
number density, i.e., we compare the Poisson point process, the
equilibrium hard-sphere liquid, and the MRJ state at the same number
density $\rho=1$ (the unit volume contains one particle on
average)~\footnote{Because the packing fraction of the equilibrium
  hard-sphere liquid is lower than for the MRJ state, the
  corresponding diameters of the spheres are different in this
  paper.}.

Because the number density is set to unity, the mean cell volume is
also one. The average surface area and integrated mean curvature of a
Voronoi cell in the MRJ state or in the equilibrium ensemble are
slightly larger than those of a Poisson Voronoi cell because the
latter is less regular, i.e., more aspherical.
The Voronoi volume fluctuations and the standard deviations of the
other Minkowski functionals are much stronger in the irregular Poisson
point process than in the hard-sphere packings, where the MRJ state
has significantly smaller Voronoi volume fluctuations than the
equilibrium hard-sphere liquid. The Minkowski functionals of a single
Voronoi cell, e.g., its volume and its surface area, are strongly
correlated, i.e., the correlation coefficients $\rho_{\mu,\nu}$ are at
least $0.9$. The numerical estimates for the Poisson Voronoi
tessellation are in agreement with the analytic values and numerical
estimates in Ref.~\cite{ChiuEtAl2013} and references therein.

\begin{figure}[t]
  \centering
  \includegraphics[width=\linewidth]{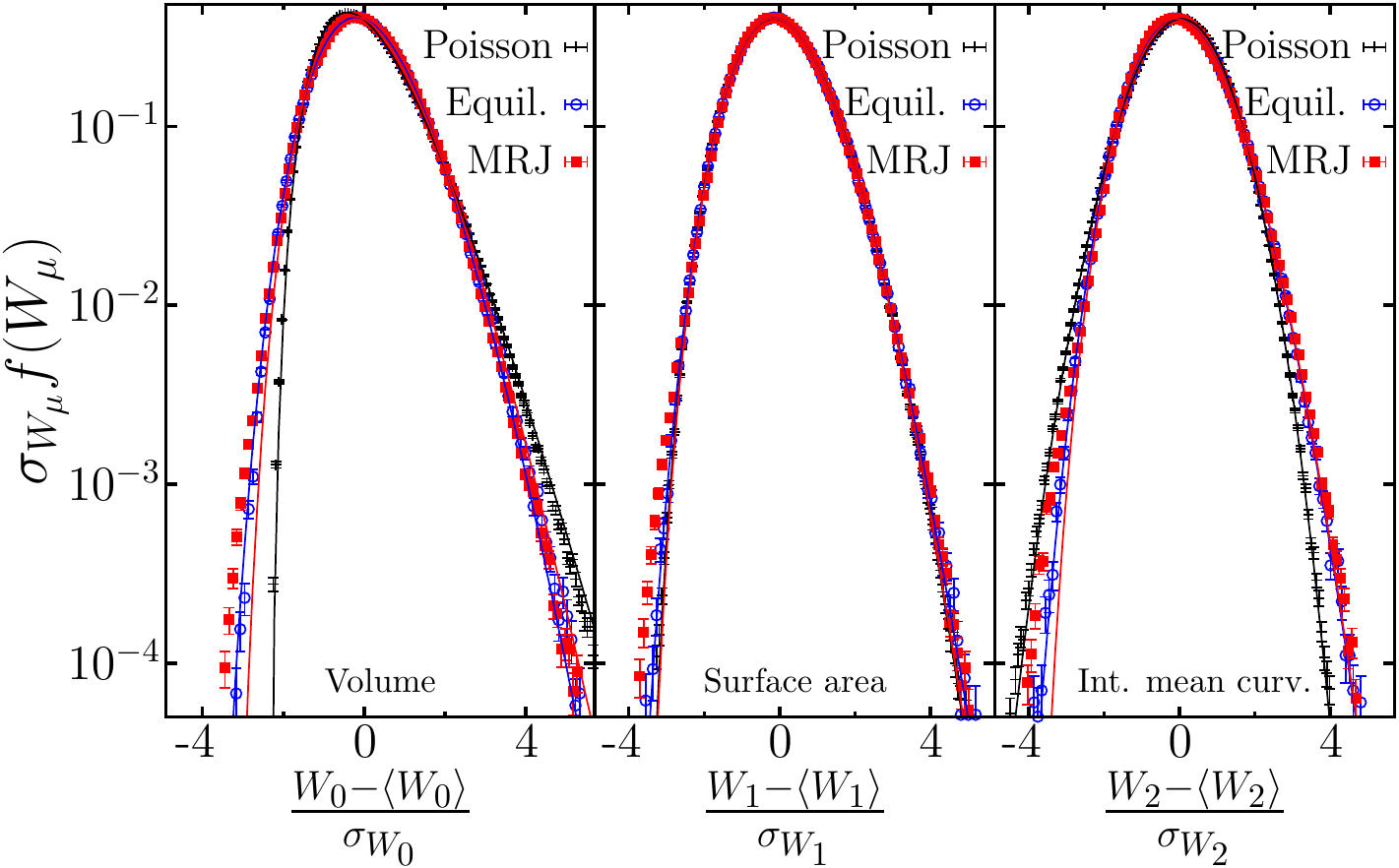}
  \caption{(Color online) Distributions of the single Minkowski functionals $W_{\mu}$
    of a three-dimensional Voronoi cell in a Poisson point process, an
    equilibrium hard-sphere system at a packing fraction $\phi =
    0.48$, and a MRJ sphere packing: $\mu=0$ volume (left), $\mu=1$
    surface area (center), and $\mu=2$ integrated mean curvature
    (right). The distributions are rescaled---like in
    Ref.~\cite{StarrEtAl2002}---by their mean $\langle W_{\mu}
    \rangle$ and their standard deviation $\sigma_{W_{\mu}}$ (see
    Table~\ref{tab:minkfunc}). The lines in the plot of the volume
    distributions are $\gamma$ distributions; generalized $\gamma$
    distributions are fitted to the distributions of the surface area
    and the integrated mean curvature.}
  \label{fig:single_MF_distributions}
\end{figure}

The high fidelity of the MRJ sphere packings produced by the TJ
algorithm allows one to study the relation between the number of
contacts of a sphere and the Minkowski functionals of its Voronoi
cell. As expected, small cells have a higher number of contacts on
average because a high local packing fraction~\footnote{The local
  packing fraction is the volume of the sphere divided by the volume
  of the Voronoi cell.} implies that there are many close
neighbors. In units of $\lambda$, the mean Voronoi volume of a
rattler, i.e., an unjammed particle, is 1.04 and that of a particle
with 11 contacts is 0.88. The mean surface area of the Voronoi cells
of rattlers and of backbone spheres with up to 11 contacts varies from
5.50 to 4.92, respectively, and the average integrated mean curvature
varies from 8.65 to 8.17, respectively.
However, because of the small difference between near contacts and
true contacts, the distributions of the Minkowski functionals for
rattlers are only slightly shifted compared to the distributions of a
typical cell. There are, for example, very small cells containing
rattlers, which is consistent with previous
findings~\cite{AtkinsonEtAl2013}.

Starr \textit{et al.}~\cite{StarrEtAl2002} and, similarly, Aste
\textit{et al.}~\cite{AsteMatteo2008} showed that by shifting the
volume distribution by its mean and rescaling with its standard
deviation, the volume distributions of many different sphere packings
collapse.
Figure~\ref{fig:single_MF_distributions} shows the rescaled
distributions of the Minkowski functionals for the Poisson point
process, the equilibrium hard-sphere liquid, and the MRJ packing. As
expected, the volume distributions of the equilibrium hard-sphere
packings and the MRJ packings are qualitatively very similar, while
the distribution of the Poisson point process deviates. The same is
true for the distribution of the mean curvatures. The distributions of
the surface area for both the MRJ and the equilibrium hard-sphere
packings are not only qualitatively similar to each other but also to
the uncorrelated Poisson point process.
So, besides the quantitative difference in the mean and the standard
deviations of the Minkowski functionals, the distributions of the
Minkowski functionals of single Voronoi volumes are qualitatively
similar for the equilibrium hard-sphere liquid and the MRJ state as
well as even partially for the Poisson point process.
The distribution of the Minkowski functionals of a single cell only
incorporates local information and is rather insensitive to global
structural features such as hyperuniformity of the MRJ
state~\cite{DonevEtAl2005, HopkinsStillingerTorquato2012}.

\begin{figure}[b]
  \centering
  \includegraphics[width=\linewidth]{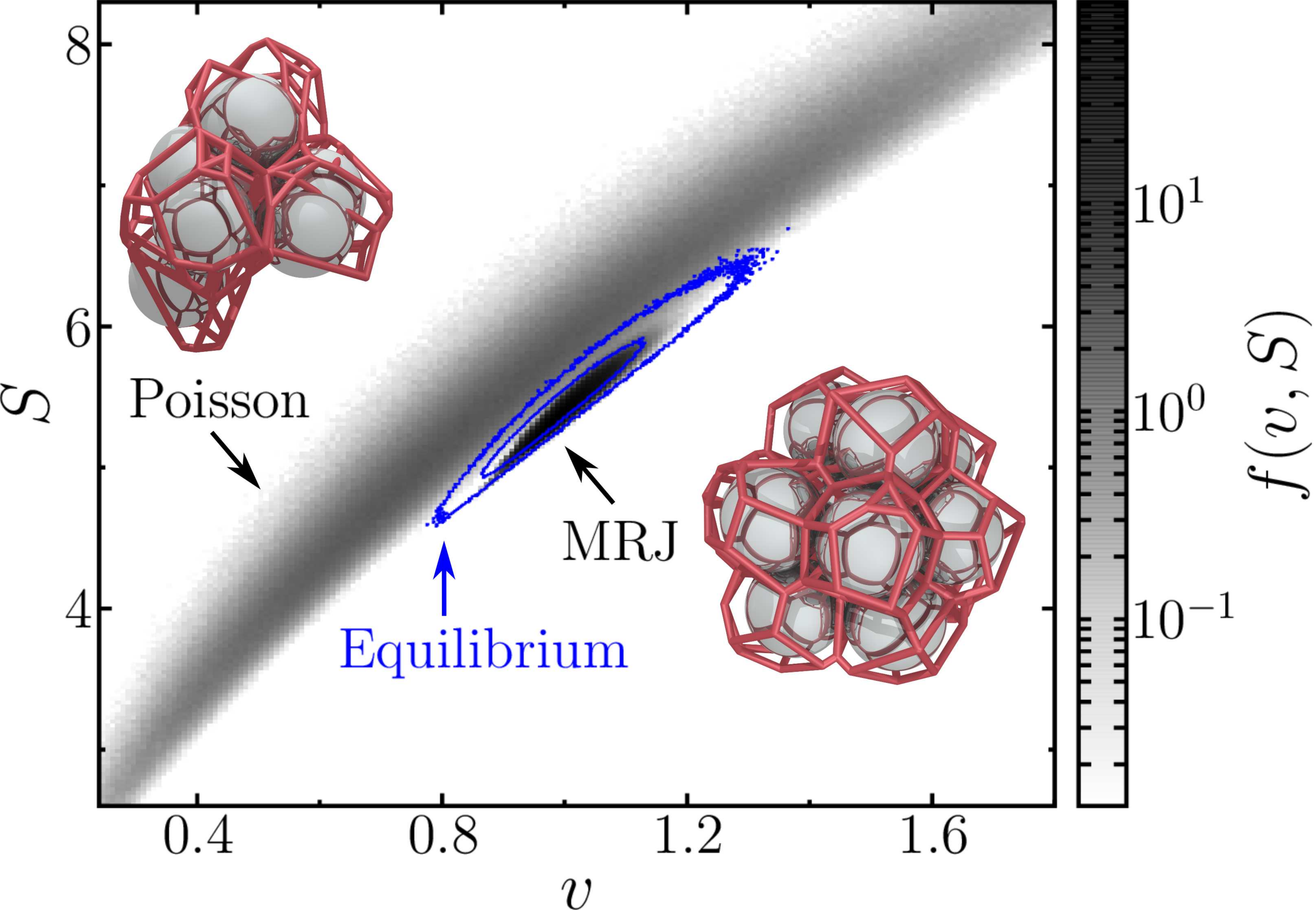}
  \caption{(Color online) Joint distribution of volume $v$ and surface area $S$ of a
    single three-dimensional Voronoi cell in a logarithmic scale for a
    Poisson point process and a MRJ sphere packing; the joint
    distribution for the equilibrium hard-sphere liquid is represented
    by the blue contour plot. The unit of length is
    $\lambda=1/\rho^{1/3}$, where $\rho$ is the number density. Samples
    of a MRJ sphere packing and an overlapping sphere packing, where
    the sphere centers follow a Poisson point process, are depicted
    together with their Voronoi diagrams.}
  \label{fig:joint_MF_distributions}
\end{figure}

Figure~\ref{fig:joint_MF_distributions} shows the joint probability
distribution of the volume and the surface area of a single Voronoi
cell in a Poisson point process, an equilibrium hard-sphere liquid,
and a MRJ sphere packing. The joint probability distributions for
the equilibrium hard-sphere liquid and the MRJ state are also
relatively similar.

Both for the Poisson point process~\cite{Pineda2004} and for many
different numerical and experimental sphere
packings~\cite{AsteMatteo2008}, the volume distribution follows well a
$\gamma$ distribution~\footnote{The volume distributions are well
  approximated by shifted $\gamma$ distributions,
  \[f(v)=\frac{1}{\Gamma(k)\theta^k}(v-v_{\mathrm{min}})^{k-1}\mathrm{e}^{-(v-v_{\mathrm{min}})/\theta}\]
  where the parameters {$k=(\langle v \rangle -
    v_{\mathrm{min}})^2/\sigma_v^2$} and {$\theta=\sigma_v^2/(\langle
    v \rangle - v_{\mathrm{min}})$} are determined by the mean
  {$\langle v \rangle$} and standard deviation $\sigma_v$ of the
  volume distribution~\cite{AsteMatteo2008}. The volume of a Voronoi
  cell of a hard-sphere packing must be greater than the volume
  $v_{\mathrm{min}}$ of a dodecahedron with the sphere touching all of
  its faces~\cite{AsteMatteo2008, TorquatoJiao2013Dodecahedron}.}.
We also find, for the volume distributions for the Poisson point process
and the equilibrium hard-sphere liquid, an excellent agreement with
$\gamma$ distributions~\footnote{The parameters $k$ of the $\gamma$
  distributions are $k= 5.6$ for the Poisson point process, $k= 24.3$
  for the equilibrium hard-sphere liquid, and $k= 13.3$ for the MRJ
  sphere packings.}.
However, for the MRJ sphere packings there is a slight but
statistically significant deviation for very small cells for which the
frequency of occurrence is too high.
The surface area and the integrated mean curvature distributions are
well approximated by generalized $\gamma$ distributions~\footnote{The
  generalized $\gamma$ distribution is
  \[\gamma_q(x):=\frac{q(x-x_{\mathrm{min}})^{q(k-1)}\mathrm{e}^{-((x-x_{\mathrm{min}})/\theta)^q}}{\Gamma(k-1+1/q)\theta^{1+q(k-1)}}\]
  where the parameter $\theta$ is fixed by the mean of the surface
  area or of the integrated mean curvature; the parameters $k$ and $q$
  are fit parameters. The minimum surface area and integrated mean
  curvature is assumed to be that of a dodecahedron. Note that Lazar
  \textit{et al.}~\cite{LazarEtAl2013} showed with very high
  statistics, using 250 000 000 cells, that the Voronoi volume
  distribution of a Poisson point process also deviates statistically
  significantly from a two-parameter $\gamma$ distribution, but is
  very well described by the generalized $\gamma$ distribution.},
which was already found for the Poisson point process by
Refs.~\cite{KapferEtAl2010, LazarEtAl2013}. However, the distributions
for the MRJ sphere packings deviate slightly but statistically
significantly from a generalized $\gamma$ distribution for cells with
small surface area or small integrated mean curvature,
respectively~\footnote{The reduced $\chi$ squared of the fits to the
  surface area distributions and integrated mean curvature
  distributions are for the equilibrium hard-sphere packings $0.94$ or
  $0.96$, respectively, and for the Poisson point process $2.40$ or
  $2.43$, respectively, but for the MRJ sphere packings, they are
  $23.2$ or $25.7$, respectively.}.

\section{Correlation Functions and Probability Density Functions of
  Minkowski Functionals}
\label{sec:definitions}

In order to quantify the global structure of the Voronoi diagram,
correlation functions of the Minkowski functionals of cells at a
distance $r$ and cell-cell probability density functions are
introduced and defined here.

\subsection{Correlation Functions of Minkowski Functionals} 
\label{sec:c0}

We define the volume-volume correlation function
$C_{00}(\vec{r}_1,\vec{r}_2)$ of the Voronoi cells of an arbitrary point
process as the correlation between the volume of two Voronoi cells
given that the corresponding centers are at the positions $\vec{r}_1$
and $\vec{r}_2$:
\begin{align}
  C_{00}(\vec{r}_1,\vec{r}_2) & := \frac{\langle v(\vec{r}_1) v(\vec{r}_2)
  \rangle - \langle
  v(\vec{r}_1)\rangle \langle v(\vec{r}_2)
  \rangle}{\sigma_{v(\vec{r}_1|\vec{r}_2)}\sigma_{v(\vec{r}_2|\vec{r}_1)}}
\label{eq:c0}
\end{align}
where $\langle . \rangle$ denotes the ensemble average given two
points at $\vec{r}_1$ and $\vec{r}_2$; and 
$\sigma_{v(\vec{r}_i|\vec{r}_j)}$ is the standard deviation of the
volume $v$ of the Voronoi cell at $\vec{r}_i$ given that there is
another point at $\vec{r}_j$. Note that because of this condition,
both the mean and the standard deviation of a single Voronoi volume
are functions of the positions $\vec{r}_1$ and $\vec{r}_2$: e.g.,
knowing that there is a point in close proximity, very large
volumes are less likely and the mean volume decreases.
For a statistically homogeneous and isotropic point process, the
volume-volume correlation is simply a radial function, which we denote
by $C_{00}(r)$, where $r=\|\vec{r}_2 -\vec{r}_1\|$.
The correlation function $C_{00}(r)~\in~[-1;1]$ measures the
correlations, both positive and negative (anticorrelations),
 between Voronoi volumes of cells given that their centers
are at a distance $r$.

The Voronoi tessellation assigns to each point a volume of its
corresponding Voronoi cell. This is a special case of a marked point
process where the constructed mark assigned to each point is
determined by the positions of the points in the neighborhood. In this
sense, the volume-volume correlation function can be seen as a special
type of a marked correlation
function~\cite{MeckeStoyan2000,ChiuEtAl2013,IllianEtAl2008}.

The volume-volume correlation function does not, in general, converge to
perfect correlation for vanishing radial distance $\lim_{r\rightarrow
  0} C_{00}(r) < 1$ because for all $r>0$ the correlation function
$C_{00}(r)$ provides the correlation of the Voronoi volumes of two
different cells with volumes $v(0)$ and $v(r)$. Because the cell is
perfectly correlated with itself, i.e., $C_{00}(0)=1$, the correlation
function $C_{00}(r)$ is discontinuous at the origin.
If there is no long-range order, the correlation function tends to
zero for infinite radial distance $\lim_{r\rightarrow \infty} C_{00}(r) =
0$.

The correlation functions of the other Minkowski functionals are
defined analogously to Eq.~\eqref{eq:c0}, replacing volume ($\mu=0$)
by surface area ($\mu=1$) or integrated mean curvature ($\mu=2$):
\begin{align}
  C_{\mu\mu}(\vec{r}_1,\vec{r}_2) & := \frac{\langle W_{\mu}(\vec{r}_1) W_{\mu}(\vec{r}_2)
  \rangle - \langle
  W_{\mu}(\vec{r}_1)\rangle \langle W_{\mu}(\vec{r}_2)
  \rangle}{\sigma_{W_{\mu}(\vec{r}_1|\vec{r}_2)}\sigma_{W_{\mu}(\vec{r}_2|\vec{r}_1)}}
  \label{eq:cnu}
\end{align}
with $\sigma_{W_{\mu}(\vec{r}_i|\vec{r}_j)}$ the standard deviation of
the Minkowski functional $W_{\mu}$ of the Voronoi cell at $\vec{r}_i$
given that there is another point at $\vec{r}_j$. For a statistically
homogeneous and isotropic point process, the correlation function of
the Minkowski functionals is again a radial function, which we denote
by $C_{\mu\mu}(r)$. In general, $C_{\mu\mu}(r)$ will be discontinuous
for $r\rightarrow 0$, as noted above for the volume-volume correlation
function.

In the Appendix, we calculate the volume-volume correlation function
analytically for the one-dimensional Poisson point process.
In Sec.~\ref{sec:3d}, we determine the correlation functions for the
three-dimensional Poisson point process, the equilibrium hard-sphere
liquid, and MRJ sphere packings.

A different type of correlation function, a pointwise Voronoi
correlation function, assigns to arbitrary points the volumes of the
Voronoi cells in which they lie~\cite{ZhaoEtAl2012}. Correlations
between Voronoi volumes have also already been studied by finding a
nonlinear scaling in the aggregate Voronoi volume fluctuations as a
function of the sample size~\cite{AsteMatteo2007}.

\subsection{Cell-Cell Probability Density Functions of the Voronoi
  Volume}
\label{sec:pdf}

The volume-volume correlation function $C_{00}(\vec{r}_2,\vec{r}_1)$
is defined conditionally on the fact that the centers of the two cells
are at $\vec{r}_1$ and $\vec{r}_2$.
The full two-point information about the Voronoi volumes is given by
the cell-cell probability density function
$p(\vec{r}_2,v,\vec{r}_1,v^*)$ of finding two points in the point
process at two arbitrary positions $\vec{r}_2$ and $\vec{r}_1$ with
associated Voronoi volumes $v$ and $v^*$, respectively.
It quantifies, for example, how likely it is to find near a point with
a small Voronoi cell another point with either a large or another
small Voronoi cell.
Integrating over the volumes yields the standard pair-correlation function,
\begin{align}
  g_2(\vec{r}_2,\vec{r}_1)=\iint\mathrm{d}v\,\mathrm{d}v^*\,\frac{p(\vec{r}_2,v,\vec{r}_1,v^*)}{\rho(\vec{r}_2)\rho(\vec{r}_1)}\, .
\end{align}
This relation clearly indicates that the Minkowski probability density function
$p(\vec{r}_2,v,\vec{r}_1,v^*)$ contains more information than $g_2(\vec{r}_2,\vec{r}_1)$.
Moreover, the volume-volume correlation function
$C_{00}(\vec{r}_2,\vec{r}_1)$ from Sec.~\ref{sec:c0} follows from
calculating the moments $\langle vv^*\rangle$, $\langle v \rangle$,
and $\langle v^* \rangle$ of
$\frac{p(\vec{r}_2,v,\vec{r}_1,v^*)}{\rho(\vec{r}_2)\rho(\vec{r}_1)g_2(\vec{r}_2,\vec{r}_1)}$
and the corresponding standard deviations $\sigma_{v}$ and
$\sigma_{v^*}$.

For a statistically homogeneous and isotropic point process, the
cell-cell probability density function is a radial function, i.e., it
only depends on the radial distance $r=|\vec{r}_2 -\vec{r}_1|$:
$p(r,v,v^*)$ is the probability density of finding two points with
Voronoi volumes $v$ and $v^*$ at a radial distance $r$.
If there is no long-range order, the cell-cell probability density
function $p(r,v,v^*)$ converges for large radii $r\rightarrow\infty$
to $\rho^2 f(v)f(v^*)$, with $\rho$ the number density and $f(v)$ the
distribution of the Voronoi volume $v$ of a single cell (see
Sec.~\ref{sec:MF_distributions}).

For a better visualization and comparison of different volumes, we
divide the cell-cell probability density function by its long-range
value; the cell-cell pair-correlation function is defined as
\begin{align}
  g_{vv}(\vec{r}_2,v,\vec{r}_1,v^*):=\frac{p(\vec{r}_2,v,\vec{r}_1,v^*)}{\rho(\vec{r}_2)f(v)
  \rho(\vec{r}_1)f(v^*)}
\end{align}
and, for a homogeneous and isotropic system,
\begin{align}
  g_{vv}(r,v,v^*):=\frac{p(r,v,v^*)}{\rho^2 f(v)f(v^*)}\, .
\end{align}
If $g_{vv}(r,v,v^*) > 1$, it is more likely to find a pair of Voronoi
cells with volumes $v$ and $v^*$ at a distance $r$ than to find them
at a large distance, i.e., uncorrelated. If $g_{vv}(r,v,v^*) < 1$, the
occurrence of a point in the point process with a Voronoi volume $v$
at a distance $r$ of another Voronoi center with a Voronoi volume
$v^*$ is suppressed. Analogous cell-cell pair-correlation functions
can be defined for the other Minkowski functionals.

We analytically calculate the cell-cell probability density function
for the one-dimensional Poisson point process in the Appendix.
In Sec.~\ref{sec:radial_prob}, we determine the cell-cell pair
correlation function for the three-dimensional Poisson point process,
the equilibrium hard-sphere liquid, and MRJ sphere packings.

\section{Correlation Functions of Minkowski functionals}
\label{sec:3d}

In order to sample the correlation functions of the Minkowski
functionals, the distances of all pairs of particles~\footnote{The
  distance of two particles is their minimum image separation distance
  if periodic boundary conditions are applied.} are computed and
assigned to a bin. For each radial distance, i.e., for each bin, the
correlation coefficient of the Minkowski functionals of the two
Voronoi cells is determined.

\begin{figure}[t]
  \centering
  \includegraphics[width=\linewidth]{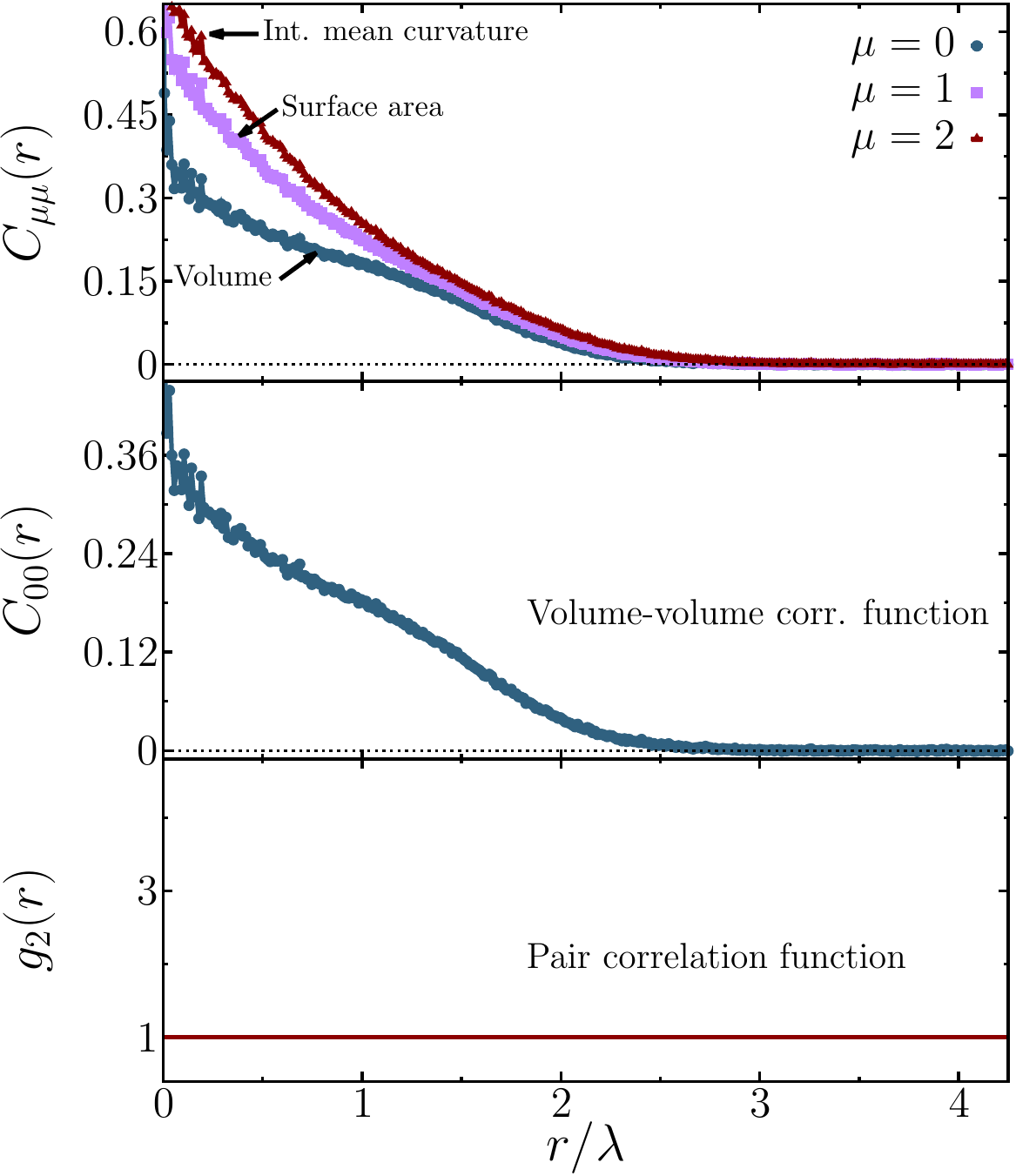}
  \caption{(Color online) Correlation functions for a
    three-dimensional \textit{Poisson point process}: pair-correlation
    function $g_2(r)$ (bottom); volume-volume correlation function
    $C_{00}(r)$ (center), which is the correlation function of the
    volumes of two Voronoi cells given that their centers are at a
    distance $r$; the mark correlation functions of the three
    different Minkowski functionals (top): $\mu=0$ volume, $\mu=1$
    surface area, and $\mu=2$ integrated mean curvature. The radial
    distance $r$ is normalized by $\lambda=1/\rho^{1/3}$, where $\rho$
    is the number density.}
  \label{fig:3d_poisson}
\end{figure}

Figures~\ref{fig:3d_poisson}--\ref{fig:3d_MRJ} compare the correlation
functions of the Minkowski functionals for the Poisson point process,
equilibrium hard-sphere liquids, and MRJ sphere packings. It is seen that these Minkowski
correlation functions contain visibly more information than
the corresponding standard pair-correlation functions, even in the case
of the Poisson point process.

\subsection{Poisson Point Process}
\label{sec:CmuPoisson}

It is evident that in the infinite-system limit, the pair-correlation
function $g_2(r)$ is a constant (unity) for the Poisson point process,
i.e., the points are completely uncorrelated. Because a Voronoi cell
is determined by the neighbors of its center, the volume will
obviously be correlated; see Fig.~\ref{fig:3d_poisson}. There are
large Voronoi volume fluctuations for the Poisson point process. Very
large cells lead to a strong correlation of the Voronoi volumes even
for distances up to four times the mean nearest-neighbor distance.
This is to be contrasted with the standard pair-correlation function
$g_2(r)$, which is trivially unity for all radial distances.

Figure~\ref{fig:3d_poisson} compares the correlation functions
$C_{\mu\mu}(r)$ for all Minkowski functionals $\mu=0,1,2$. All
functionals have approximately the same correlation length. For
$r\rightarrow 0$, the surface areas are more strongly correlated than
the volumes because at small radial distances, the cells will most
likely share a face. In the Appendix, we calculate $C_{00}(r)$
analytically for the one-dimensional Poisson point process.

\subsection{Equilibrium Hard-Sphere Liquid}

\begin{figure}[t]
  \centering
  \includegraphics[width=\linewidth]{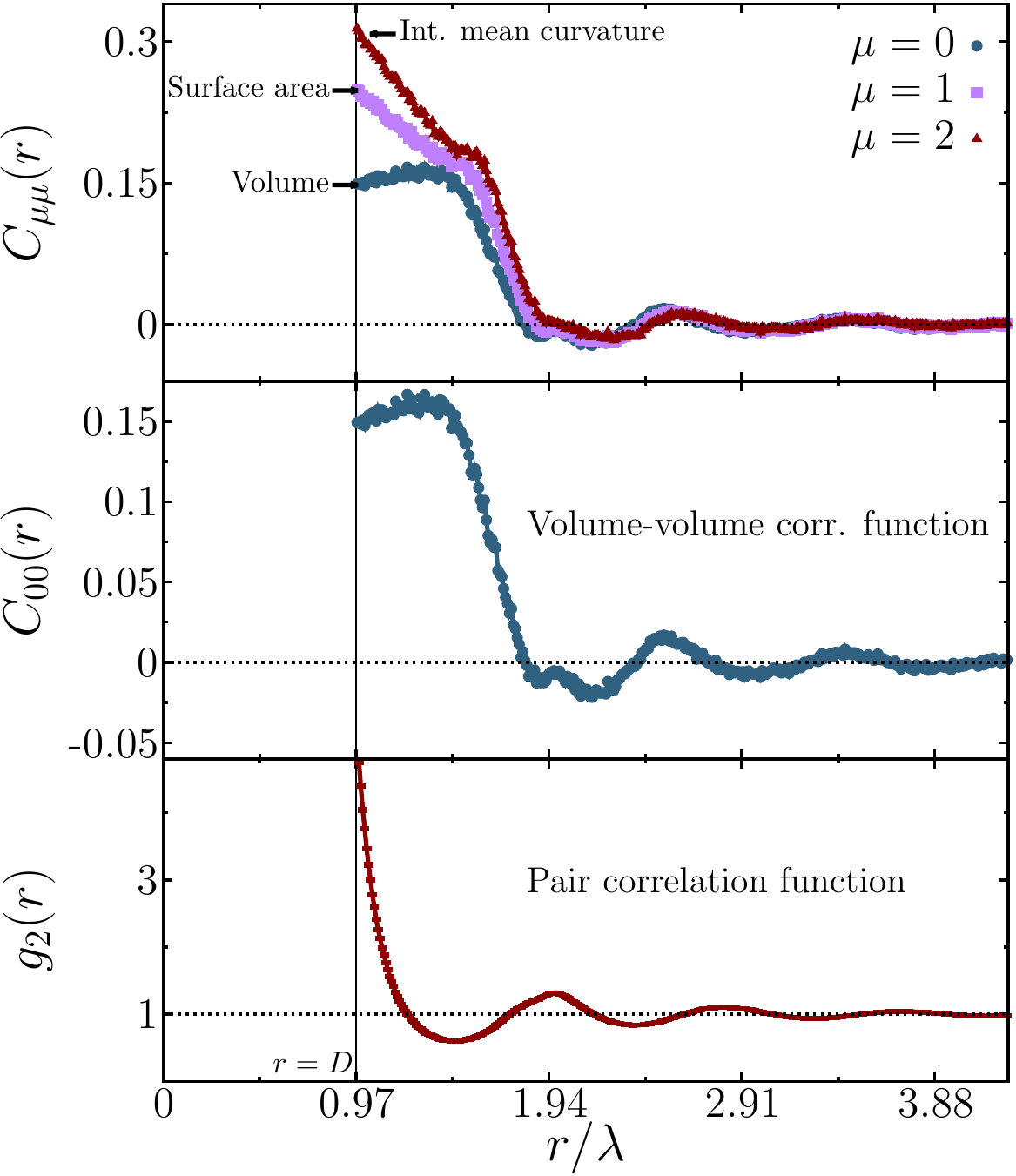}
  \caption{(Color online) Correlation functions for \textit{equilibrium hard spheres} with
    diameter $D$ at a packing fraction $\phi=0.48$; for details, see
    Fig.~\ref{fig:3d_poisson}.}
  \label{fig:3d_Equilibrium}
\end{figure}

Figure~\ref{fig:3d_Equilibrium} shows the pair-correlation function
and the correlation functions of the Minkowski functionals for
equilibrium hard-sphere liquid configurations at a packing fraction $\phi=0.48$. Because
the hard spheres are impenetrable, the correlation functions of the
Minkowski functionals are only defined for radial distances larger or
equal to the diameter $D$ of a sphere; in this case,
$D\approx0.97\,\lambda$.

\begin{figure}[t]
  \centering
  \includegraphics[width=\linewidth]{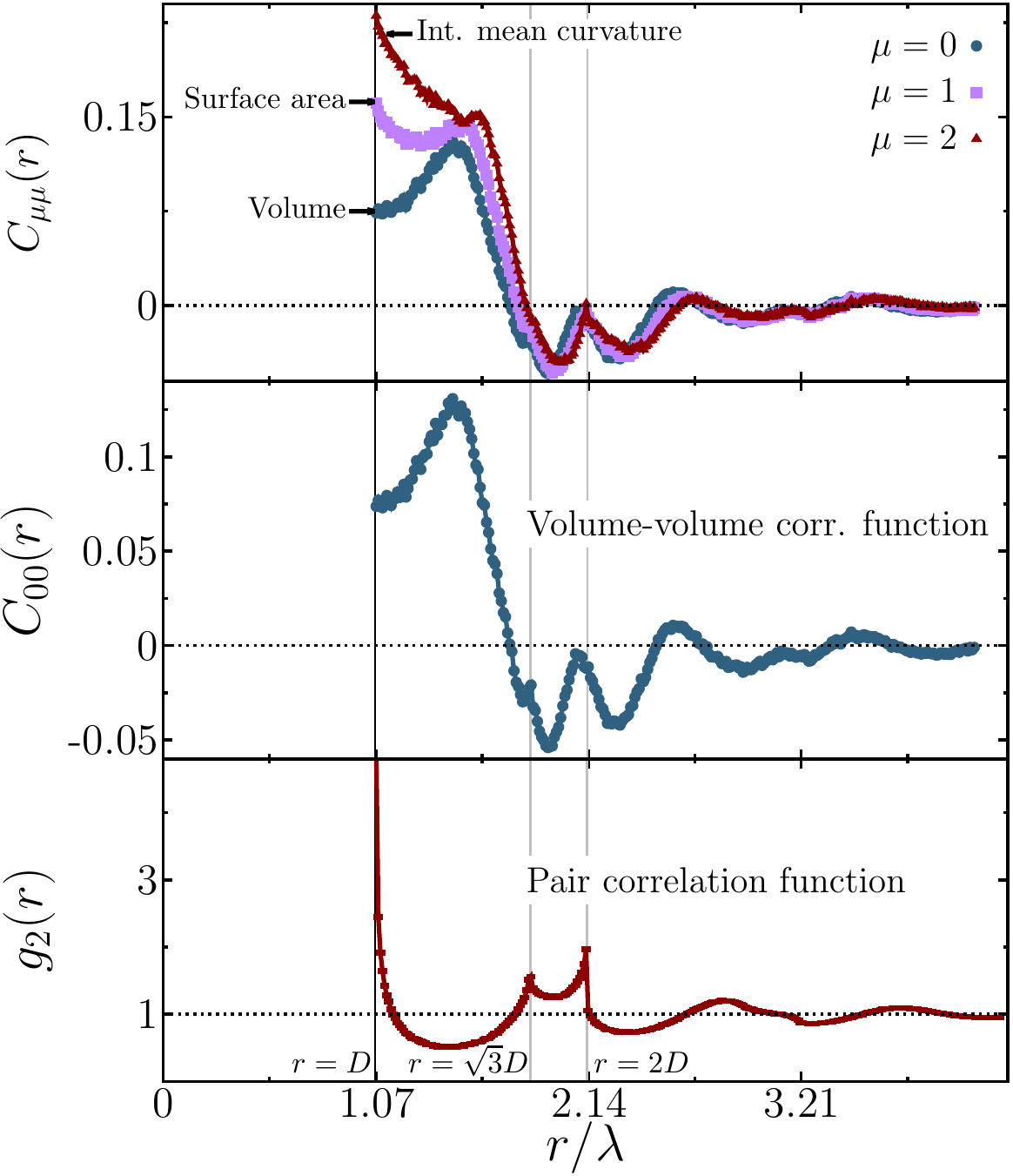}
  \caption{(Color online) Correlation functions for \textit{MRJ sphere packings} of
    spheres with diameter $D$; the average packing fraction is
    $\phi\approx0.64$. For details, see Fig.~\ref{fig:3d_poisson}. The
    pair-correlation function $g_2(r)$ is in agreement with previous
    results for MRJ sphere packings~\cite{AtkinsonEtAl2013}.}
  \label{fig:3d_MRJ}
\end{figure}

There is a strong correlation of the Voronoi volumes of spheres that
are in near contact because the Voronoi neighbors are correlated by
construction of the Voronoi diagram. However, the maximum correlation
is reached for noncontacting spheres at $r \approx 1.3\,\lambda$; a
large cell has many neighbors and a Voronoi neighbor with a sphere not
in contact will be, on average, larger than another neighbor cell with a
contacting sphere.

Between $1.8\,\lambda$ and $2.4\,\lambda$, there is a double peak of
anticorrelation and, for larger radial distances, there is an
oscillating anticorrelation and correlation similar to the pair
correlation function $g_2$, but nearly inverted. The correlation
length of the Voronoi volumes in the hard-sphere liquid is larger than
in the uncorrelated Poisson point process, where the correlation was
only due to the large Voronoi volume fluctuations.

At the top of Fig.~\ref{fig:3d_Equilibrium}, the correlation functions
of the other Minkowski functionals are compared. Similar to the
Poisson case, the integrated mean curvature is more strongly
correlated at contact $r=D$ than the surface area which, in turn, is
more strongly correlated than the volume. There is no double
anticorrelation peak in the integrated mean curvature. For large
radii, the correlation functions are shifted against each other
despite the strong correlation of the different functionals for a
single Voronoi cell. The surface area and the integrated mean
curvature are slightly less (anti)correlated.

\subsection{MRJ Sphere Packings}

The pair-correlation function $g_2(r)$ and the correlation functions
of the Minkowski functionals of the MRJ sphere packings are shown in
Fig.~\ref{fig:3d_MRJ}. The diameter of the spheres in the MRJ sphere
packings is $D\approx1.07\lambda$. The most striking differences in
the pair correlation of the jammed packings to the equilibrium
packings are the two discontinuities at $r=\sqrt{3}D$ and $r=2D$, the
split-second peak, which corresponds to configurations of two
edge-sharing equilateral and coplanar triangles ($r=\sqrt{3}D$) or a
linear chain of three particles ($r=2D$),
respectively~\cite{DonevEtAl2005PRE}. There is also a significant
(seemingly nonanalytic) feature of the volume-volume correlation
function $C_{00}(r)$ at $r=\sqrt{3}D$: a dip in the
anticorrelation. However, at $r=2D$, the feature is statistically
insignificant.
At least two double anticorrelation peaks are clearly resolved.

The most important qualitative differences in the volume-volume
correlation function are the much stronger anticorrelations in the
MRJ packings compared to the equilibrium packings. The correlation
with the nearest neighbors is weaker and the first anticorrelation
double peak is more than twice as strong as for the equilibrium
hard-sphere packings. The MRJ sphere packings are
hyperuniform~\cite{DonevEtAl2005, HopkinsStillingerTorquato2012},
i.e., large-scale density fluctuations are suppressed. Therefore,
strong Voronoi volume anticorrelations are necessary such that
Voronoi cells with a high local packing fraction are accompanied by
cells with rather low packing fractions, and vice versa.

Another difference between MRJ and equilibrium packings is a
stronger shift of the correlation functions of the other Minkowski
functionals. For the MRJ packings, there are radial distances,
e.g., $r=2.51\,\lambda$, at which the integrated mean curvatures are
anti-correlated [$C_{22}(2.51\,\lambda) < 0$] but the volumes are
correlated [$C_{00}(2.51\,\lambda) > 0$], and vice versa.

So, in contrast to the local Voronoi analysis, the global Voronoi
analysis of the MRJ packing reveals qualitative structural differences
to the equilibrium hard-sphere \mbox{liquid}.

\begin{figure}[t]
  \centering
  \includegraphics[width=\linewidth]{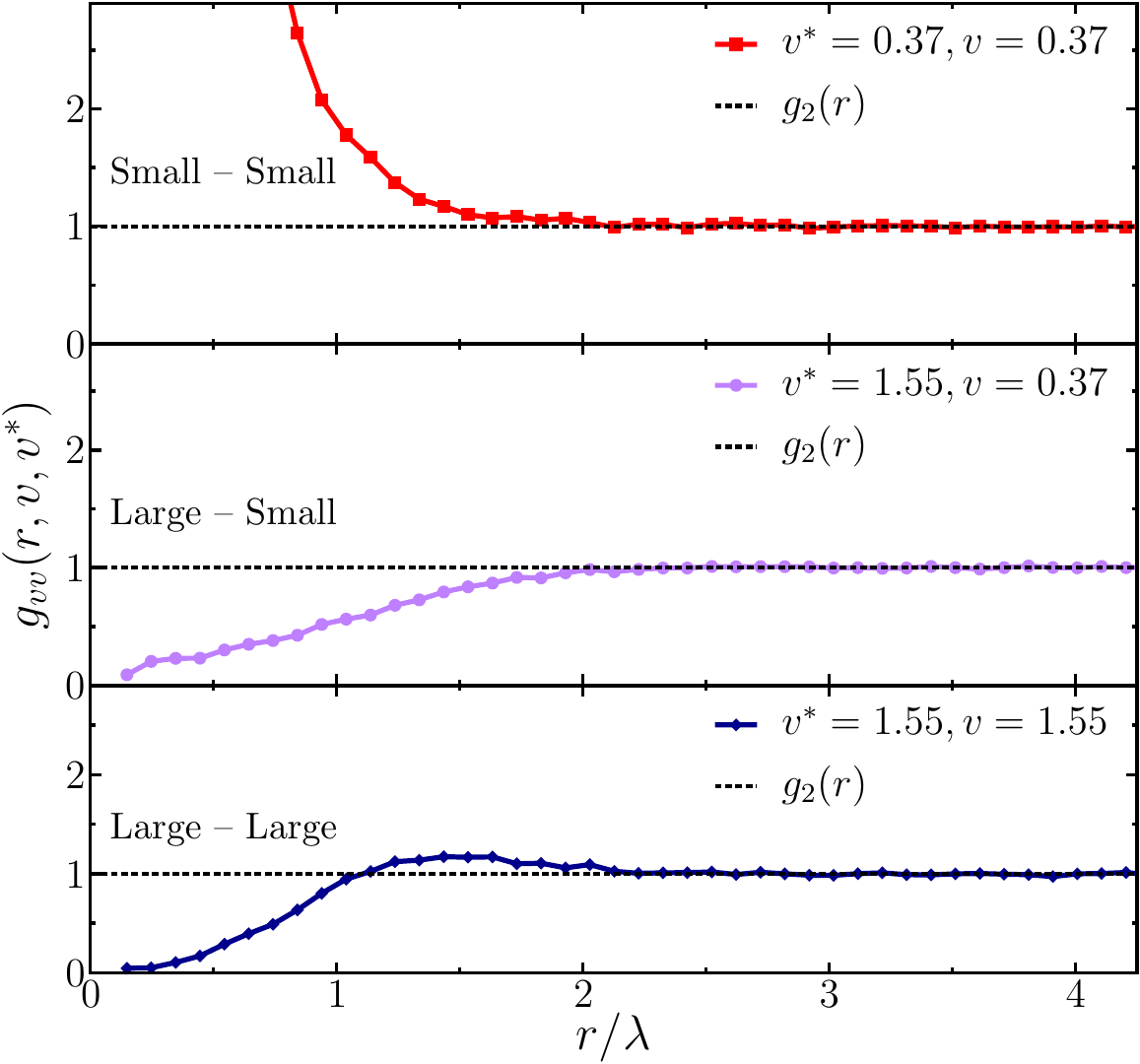}
  \caption{(Color online) The cell-cell pair-correlation function $g_{vv}(r,v,v^*)$
    for a three-dimensional \textit{Poisson point process} for either
    two large cells (bottom), a large and a small cell (center), or
    two small cells (top). As examples of large or small cells, the
    volumes were chosen such that their probability density is equal
    to $1/3$ of the maximum of the volume distribution; see
    Table~\ref{tab:minkfunc} and
    Fig.~\ref{fig:single_MF_distributions}. The curves are compared to
    the standard pair-correlation function $g_2(r)$ (dashed black
    line), which is trivially unity for the uncorrelated Poisson
    point process. The radial distance $r$ is normalized by
    $\lambda=1/\rho^{1/3}$, where $\rho$ is the number density.}
  \label{fig:pdf_poisson}
\end{figure}

\section{Cell-Cell Probability Density Functions}
\label{sec:radial_prob}

\begin{figure}[t]
  \centering
  \includegraphics[width=\linewidth]{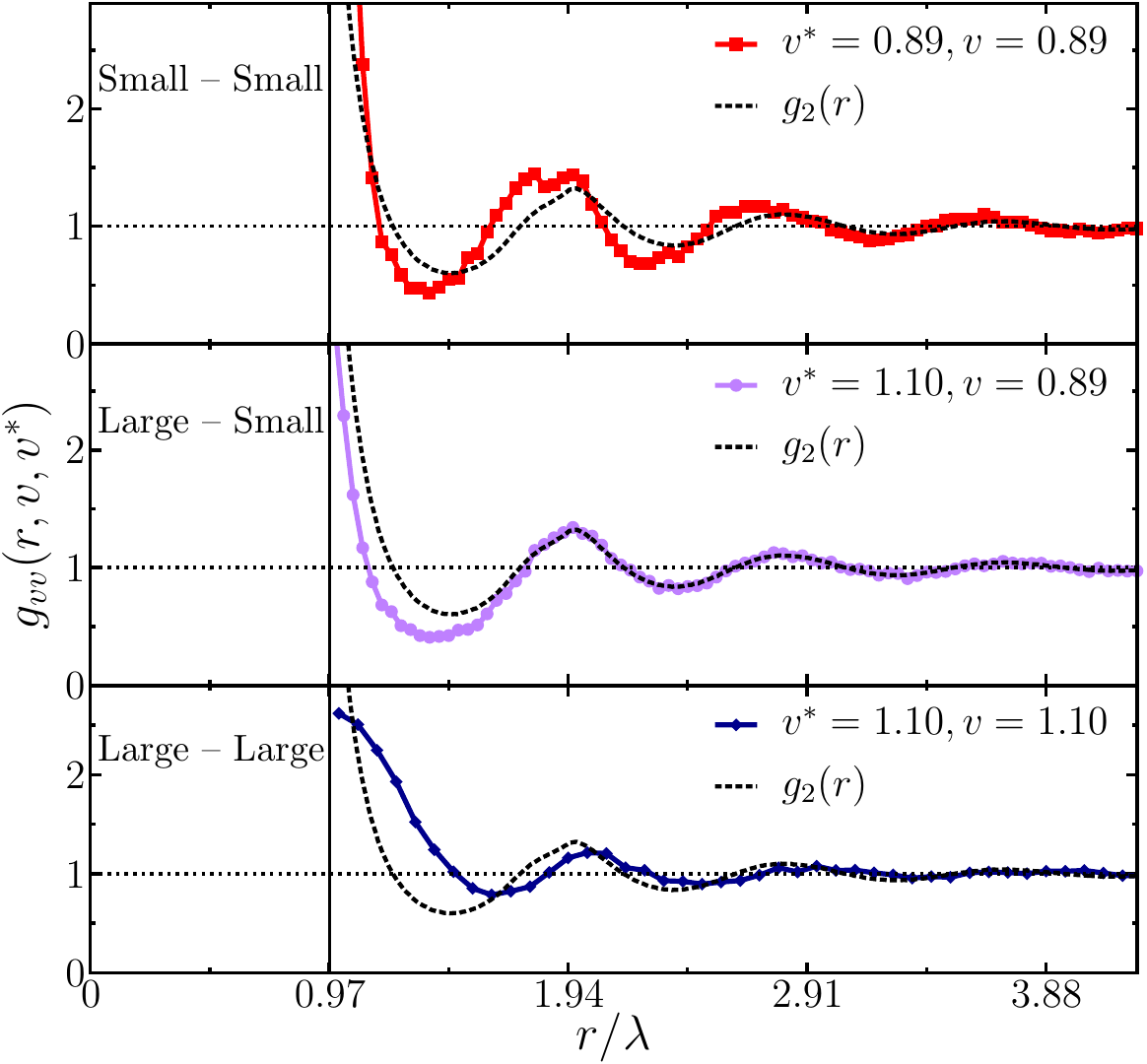}
  \caption{(Color online) The cell-cell pair-correlation function $g_{vv}(r,v,v^*)$ for
    \textit{equilibrium hard spheres} at a global packing fraction $\phi=0.48$;
    for details, see Fig.~\ref{fig:pdf_poisson}.}
  \label{fig:pdf_equilibrium}
\end{figure}

\begin{figure}[t]
  \centering
  \includegraphics[width=\linewidth]{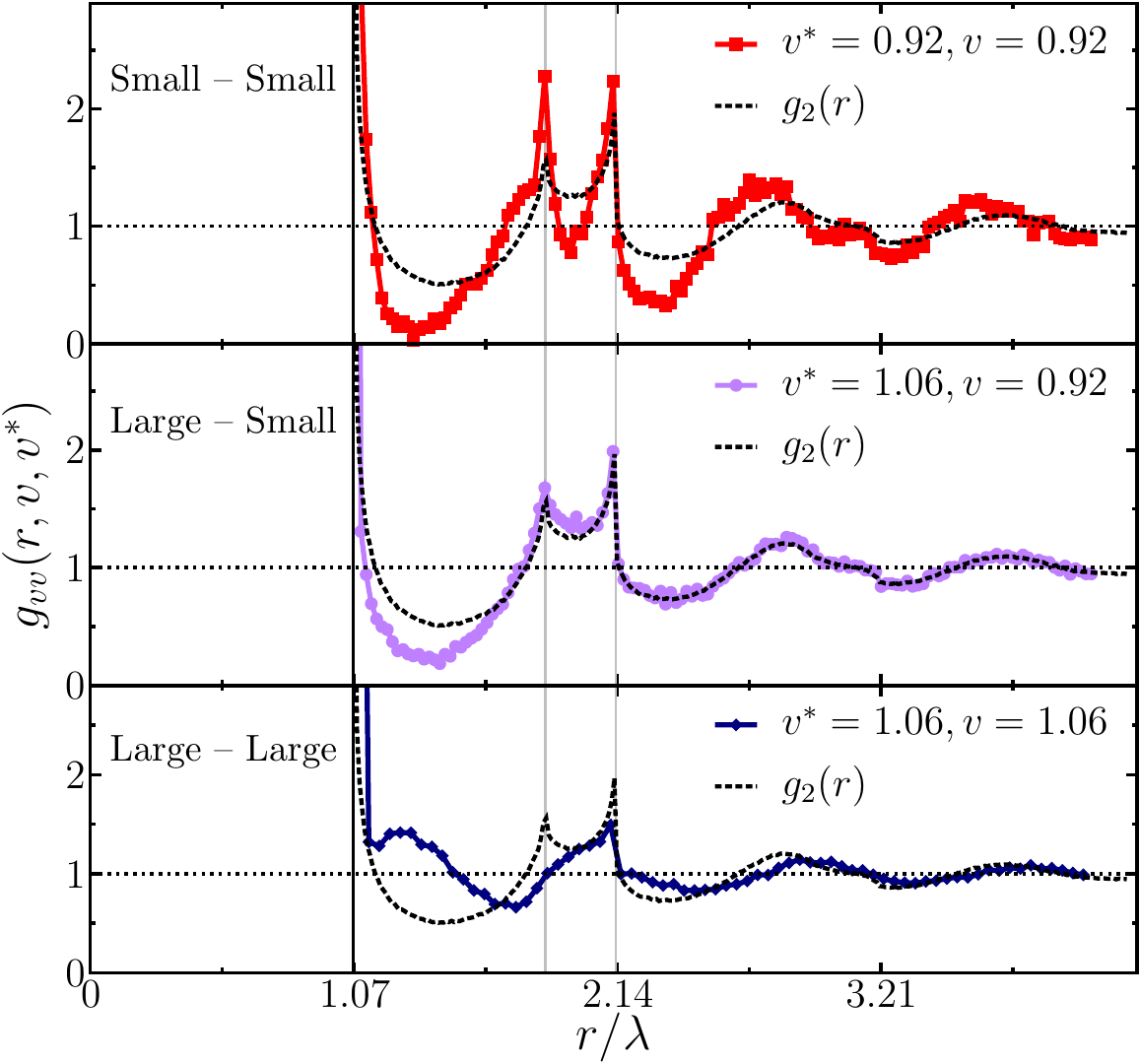}
  \caption{(Color online) The cell-cell pair-correlation function $g_{vv}(r,v,v^*)$ of the
    \textit{MRJ sphere packings}; the average packing fraction is
    $\phi\approx0.64$; for details, see Fig.~\ref{fig:pdf_poisson}.}
  \label{fig:pdf_mrj}
\end{figure}

The sampling of the cell-cell probability density function
$p(r,v,v^*)$ is very similar to that of the pair-correlation function
$g_2$ (see, e.g., Ref.~\cite{Torquato2002}); only an additional
binning with respect to the Voronoi volumes is needed.
Figures~\ref{fig:pdf_poisson}--\ref{fig:pdf_mrj} show the cell-cell
pair-correlation function $g_{vv}=p(r,v,v^*)/(\rho^2 f(v)f(v^*))$ for
exemplary large or small cell volumes $v,v^*$ in the three-dimensional
Poisson point process, in an ensemble of equilibrium hard spheres, or
in the MRJ sphere packings. As examples of large or small cells, the
volumes were chosen such that their probability density is equal to
$1/3$ of the maximum of the volume distribution; see
Table~\ref{tab:minkfunc} and Fig.~\ref{fig:single_MF_distributions}.

For the Poisson point process, the small cells are strongly correlated
at short distances because, by construction, there must be points at
close distances, and the neighbor cells of a small Voronoi cell are
more likely to be small as well. However, the probability of finding a
point with either a corresponding large or a small cell at a short
radial distance of the center of a large cell is strongly suppressed
because it is unlikely for the center of a large cell to have close
neighbors. At intermediate distances, two large cells are correlated,
as expected, because of the Voronoi construction.

In the equilibrium hard-sphere liquid, the large cells at near contact
are less correlated than the small cells. However, at slightly larger
distances, where $g_2$ shows anticorrelation and the small cells are
even more strongly anti-correlated, the large cells are positively
correlated. For distances larger than twice the diameter, the
cell-cell pair-correlation function for a large and a small cell is
equal to the standard pair-correlation function within statistical
significance. However, both the cell-cell pair-correlation functions
of finding two short or of finding two large cells at large radial
distance $r$ are shifted compared to the standard pair-correlation
function.

These features can also be found in the MRJ sphere packings. Moreover,
the anticorrelations of two small cells are much stronger. The
split-second peak even separates in two stronger peaks with
anticorrelation in between, where the standard pair-correlation
function shows positive correlation. In contrast to this, the peak at
$r=\sqrt{3}D$ completely vanishes for two large cells
$g_{vv}(\sqrt{3}D,1.06,1.06)=1$, and the peak at $r=2D$ is
significantly weaker.

\section{Conclusions and Discussions}
\label{sec:Conclusion}

We have characterized the structure of MRJ sphere packings by computing
the Minkowski functionals, i.e., the volume, the surface area, and the
integrated mean curvature, of the associated Voronoi cells.
The local analysis, i.e., the probability distribution of the
Minkowski functionals of a single Voronoi cell, provides qualitatively
similar results for the equilibrium hard-sphere liquid and the MRJ
packings and partly even for the uncorrelated Poisson point process.

In order to study the global structure of the Voronoi cells, we have
improved upon this analysis by introducing the correlation functions
$C_{\mu\mu}(r)$ of the Minkowski functionals and the cell-cell
probability density function $p(r,v,v^*)$.
The correlation function $C_{\mu\mu}(r)$ measures the correlation of
the Minkowski functionals $W_{\mu}$ of two Voronoi cells given that
the corresponding centers are at a distance $r$.
The cell-cell probability density function $p(r,v,v^*)$ also
incorporates the probability that there are two particles at a
distance $r$. For an easier interpretation and better visualization,
we have defined the dimensionless cell-cell pair-correlation function
$g_{vv}=p(r,v,v^*)/(\rho^2 f(v) f(v^*))$, where $f(v)$ is the
probability of the Voronoi volume $v$.
The generalization of the pair-correlation to the cell-cell pair
correlations provides powerful theoretical and computational tools to
characterize the complex local geometries that arise in jammed
disordered sphere packings.

Because the faces of a Voronoi cell are bisections between a point in
the point process (whether a packing or not) and its neighbors and,
moreover, because Voronoi neighbors share a face and edges, the
Minkowski functionals of neighboring Voronoi cells are correlated by
construction. This leads to a large correlation length for the Voronoi
cells in a Poisson point process because of large Voronoi volume
fluctuations.
In the equilibrium hard-sphere liquid and MRJ sphere packings, there
are correlations and anticorrelations.
In contrast to the qualitatively similar local Voronoi structure, the
global Voronoi structure of the MRJ hard-sphere packings is
qualitatively quite different from that of an equilibrium hard-sphere
liquid. We find strong Voronoi volume anticorrelations, which is
consistent with previous findings that MRJ sphere packings are
hyperuniform~\cite{DonevEtAl2005, HopkinsStillingerTorquato2012},
i.e., large-scale density fluctuations are suppressed.
MRJ sphere packings are prototypical glasses in that they have no
long-range order but they are perfectly rigid, i.e., the elastic
moduli are unbounded~\cite{TorquatoDonevStillinger2003,
  TorquatoStillinger2010RevModPhys, MarcotteEtAl2013}.
The global analysis introduced here shows the difference in the
structure of the Voronoi cells of the MRJ state and those of a
hard-sphere liquid, which further indicates that the structure of a
glass is not that of a ``frozen liquid''~\cite{TruskettEtAl2000,
  HopkinsStillingerTorquato2012, MarcotteEtAl2013}.

An already known distinct structural difference between the
hyperuniform MRJ sphere packings and equilibrium hard-sphere liquids
is that while in the equilibrium packing the total pair-correlation
function $h(r)=g_2(r)-1$ is exponentially damped, the total
correlation function of the MRJ state has a negative algebraic
power-law tail~\cite{DonevEtAl2005, HopkinsStillingerTorquato2012,
  MarcotteEtAl2013}.
It is an interesting question as to whether the asymptotic behaviors
of the correlation function of the Minkowski functionals
$C_{\mu\mu}(r)$ or the radial cell-cell correlation functions
$g_{vv}(r,v,v^*)$ are different for the MRJ state and the hard-sphere
liquid.
However, a direct observation of the power-law tail has, so far,
not been possible~\cite{DonevEtAl2005, HopkinsStillingerTorquato2012,
  MarcotteEtAl2013}; much larger systems are needed but are not available
at the moment.
Still, the global characteristics $C_{\mu\mu}(r)$ and
$g_{vv}(r,v,v^*)$, introduced in the present paper, allow for an
investigation of the underlying geometrical reasons for the negative
algebraic tail in the total pair-correlation function: the suppressed
clustering of regions with low and high local packing
fractions~\cite{HopkinsStillingerTorquato2012}.

Moreover, they also allow for a quantification of the global structure
of other cellular structures, e.g., foams, where the centers of mass
of the single cells can be used as centers of the cells instead of the
Voronoi centers used here.

A frequently discussed question is whether or not there are local
icosahedral configurations in jammed packings~\cite{Finney1976,
  ClarkeJonsson1993, AsteEtAl2005, KuritaWeeks2010, KapferEtAl2012},
i.e., a central sphere with 12 spheres in contact where the
centers of the touching spheres form a regular icosahedron. The
Voronoi cell of the central sphere in such an icosahedron is a regular
dodecahedron, which has the maximum possible local packing fraction
($\approx 0.76$). There is growing evidence, that there are no regular
icosahedral arrangements in hard-sphere packings, e.g., see
Refs.~\cite{AsteEtAl2005, DonevEtAl2005PRE, Kapfer2011}. Indeed, we
find in our MRJ sphere packings no regular and hardly any nearly
regular dodecahedral Voronoi cells.
All spheres out of more than two million have less than 12
contacts. There are local packing fractions up to $0.75$, but only
$4.2\times 10^{-5}$ of all cells have a local packing fraction greater
than 0.74.

In a preliminary approach to look for possibly strongly distorted
dodecahedral Voronoi cells in the MRJ sphere packings, we examined the
topology of the Voronoi polyhedra, i.e., the number of faces and the
corresponding types of polygons, following Refs.~\cite{Barnette1969,
  Finney1976, LazarEtAl2012}. In a compact notation, the topology of a
polyhedron is given by the so-called $p$ vector ($n_3$ $n_4$ $n_5$
$n_6$), where $n_3$ is the number of triangles, $n_4$ of
quadrilaterals, $n_5$ of pentagons, and $n_6$ of hexagons. The
dodecahedron is formed by 12 pentagons, i.e., its topology is
denoted by (0 0 12 0).
Although these polyhedron characteristics are discontinuous and
inadequately metric for definite conclusions~\cite{Finney1981}, they
can provide a first insight into whether there could be a significant
number of distorted dodecahedra.
In the MRJ sphere packings, 1.1\,\% of all cells have the topology of
a dodecahedron (0 0 12 0)~\footnote{In the equilibrium hard-sphere
  liquid, 0.45\,\% of all cells are distorted dodecahedra; in a Poisson
  Voronoi tessellation, the fraction is less than $2\times 10^{-5}$.}.
The average local packing fraction of those distorted dodecahedra is
0.69 and is thus significantly greater than the total mean local packing
fraction which is 0.64. However, only 0.4\,\% of the distorted
dodecahedra have a local packing fraction greater than 0.74. The
distorted dodecahedra also have a higher average number of contacts,
$\approx 7$, compared to the typical cell, $\approx 6$, but as stated
above there is not a single sphere with 12 contacts in this high-quality MRJ data.
There are 25 other topologies in the Voronoi diagram of the MRJ sphere
packings that occur more frequently than the dodecahedron. With
5.2\,\% of all cells, the most likely topology is (0 3 6 5). However,
by adding one or two faces, the dodecahedron can transform to the
following polyhedra~\cite{Finney1976}: 1.1\,\% of all cells in the MRJ
sphere packings are (1 0 9 3), 3.1\,\% are (0 1 10 2), and 4.4\,\% are
(0 2 8 4). The latter is the second most common type in the MRJ sphere
packings.
So, while we find no regular icosahedral configurations in the MRJ
sphere packings, the preliminary topological analysis indicates that
more detailed studies of probably strongly distorted icosahedra could
be interesting.
For example, also in metallic glasses, significant numbers of
distorted icosahedra have been found~\cite{LeeEtAl2004,
  HirataEtAl2013}.

In the second paper of this series, we will further investigate the
global structure of the MRJ sphere packings by looking at density
fluctuations, the pore-size distribution, and the two-point correlation
functions.

\begin{acknowledgments}
  M.K. thanks Klaus Mecke for continuous support and guidance and
  valuable discussions and advice. We thank Steven Atkinson, Sebastian
  Kapfer, Adil Mughal, and Gerd Schr{\" o}der-Turk for valuable
  discussions and suggestions. This work was supported in part by the
  National Science Foundation under Grants No. DMR-0820341 and
  No. DMS-1211087. We also thank the German Science Foundation (DFG)
  for Grants No. ME1361/11 ``Random Fields'' and ME1361/12 ``Tensor
  Valuations'' awarded as part of the DFG-Forschergruppe ``Geometry
  and Physics of Spatial Random Systems.''
\end{acknowledgments}

\appendix

\section{Correlation Functions of the Voronoi cells in the
  One-Dimensional Poisson Point Process}
\label{sec:1d}

As an introductory example of the global Voronoi statistics introduced
in this paper, we analytically calculate the correlation functions of
the Voronoi cells in the one-dimensional Poisson point process.
Therefore, we use the probability density functions $H_p(n_l)$ and
$H_p(n_r)$ of the nearest neighbor on the left-hand side at a distance
$n_l$ or on the right-hand side at a distance $n_r$,
respectively.

Two points $x_1$ and $x_2$ at a distance $r$ are given. Without loss
of generality, we assume in the following $x_1=0$ and $x_2=r$. The
nearest-neighbor probability density functions of $x_1$ are $H_p(n_l)
= \rho\mathrm{e}^{-\rho n_l}$ and $H_p(n_r) = \rho\mathrm{e}^{-\rho
  n_r}\theta(r-n_r)+\mathrm{e}^{-\rho r}\delta(r-n_r)$, where
$\theta(x)$ is the Heaviside step function and $\delta(x)$ is the Dirac
delta distribution. For $x_2$, the distributions for the right- and the
left-hand side simply exchange. The probability distribution
$f(v^*|r)$ of the volume $v^*$ of the cell corresponding to $x_1=0$ is
given by the average of $\delta(\frac{n_r+n_l}{2}-v^*)$:
\begin{align}
  f(v^*|r) &= \begin{cases}
    4v^*\rho^2\mathrm{e}^{-2v^*\rho} &\mbox{if }v^*<\frac{r}{2}\\
    2\rho(r\rho+1)\mathrm{e}^{-2v^*\rho} &\mbox{if }v^*\geq\frac{r}{2}\, .
  \end{cases}
  \label{eq:cond_prob}
\end{align}

Given a volume $v^*$ of the cell corresponding to $x_1$: If
$v^*<{r}/{2}$, there will be at least one additional point $y$ between
$x_1$ and $x_2$. Its distance $z$ to $x_2$ is uniformly
distributed between $r-2v^*$ and $r$. With $h(z|r,v^*)$ denoting
the probability density function of this distance, the conditional
probability distribution of the volume $v$ of the cell corresponding
to $x_2$ is given by
\begin{align}
  f(v|r,v^*) &= \int_{0}^{r}\mathrm{d}z\,h(z|r,v^*)\cdot f(v|z)
  \label{eq:cond_cond_prob}
\end{align}
with $f(v|z)$ from Eq.~\eqref{eq:cond_prob}.
A case-by-case analysis for differently large $v$ compared to
$r$ and $v^*$ is needed. If $v<\frac{r}{2}-v^*$:
\begin{align}
  f(v|r,v^*) &= 4v\rho^2\mathrm{e}^{-2v\rho}\,.
  \label{eq:asymptotics}
\end{align}
If $\frac{r}{2}-v^*<v<\frac{r}{2}$, then
\begin{equation}
  \begin{aligned}
    f(v|r,v^*) = &\frac{\rho\mathrm{e}^{-2v\rho}}{2v^*}\left[4v(r-v)\rho-(r-2v^*)^2\rho\right.\\
    &\left.+4(v+v^*)-2r\right]\,.
  \end{aligned} 
\end{equation}
If $v>\frac{r}{2}$, then
\begin{align}
  f(v|r,v^*) &= 2\rho\mathrm{e}^{-2v\rho} \left[1+\rho(r-v^*) \right]\,.
\end{align}

\begin{figure}[t]
  \centering
  \includegraphics[width=\linewidth]{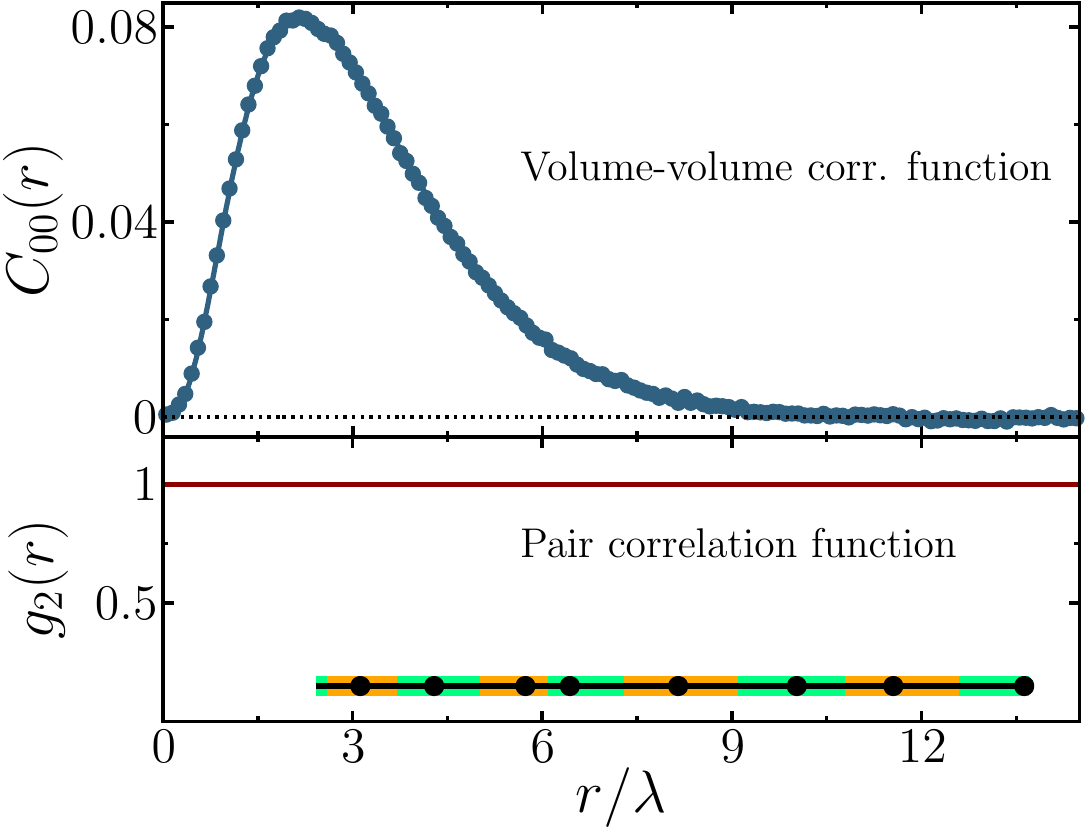}
  \caption{(Color online) Correlation functions for the
    one-dimensional Poisson point process: pair-correlation function
    $g_2(r)$ (bottom); volume-volume correlation function $C_{00}(r)$
    (top). The distance $r$ is scaled (in this one-dimensional
    example) by the inverse of the number density $\lambda=1/\rho$. An
    example of a one-dimensional Voronoi diagram of a Poisson point
    process is depicted.}
  \label{fig:1d_poisson}
\end{figure}

If $v^*>{r}/{2}$, there will be at least one additional point $y$
between $x_1$ and $x_2$ with probability $r\rho/(r\rho+1)$ and with
probability $1/(r\rho+1)$ the points $x_1$ and $x_2$ are nearest
neighbors. In the first case, the conditional probability distribution
of the volume $v$ is given by Eq.~\eqref{eq:cond_cond_prob}, but the
distance $z$ is now uniformly distributed between $0$ and $r$.
If $v<\frac{r}{2}$, then
\begin{align}
  f(v|r,v^*) &= \frac{4v\rho\mathrm{e}^{-2v\rho}}{r}((r-v)\rho+1)\,.
\end{align}
If $v>\frac{r}{2}$, then
\begin{align}
  f(v|r,v^*) &= \rho\mathrm{e}^{-2v\rho}(r\rho+2)\,.
\end{align}
In the second case, where there is no point between $x_1$ and $x_2$,
the volume $v$ is at least $r/2$ and completely determined by the
nearest neighbor of $x_2$ on the right-hand side,
\begin{align}
  f(v|r,v^*)=2\rho\mathrm{e}^{r\rho}\mathrm{e}^{-2v\rho}\,.
\end{align}

The cell-cell probability density function $p(r,v,v^*)$ from
Sec.~\ref{sec:pdf} is then given by
\begin{align}
  p(r,v,v^*) = \rho^2\cdot f(v^*|r)\cdot f(v|r,v^*)\, .
\end{align}
From Eqs.~\eqref{eq:cond_prob} and \eqref{eq:asymptotics} follows the
asymptotic behavior of $p(r,v,v^*)\xrightarrow{\: r \to \infty \:}
\rho^2 f(v)f(v^*)$.

As described in Sec.~\ref{sec:pdf}, the volume-volume correlation
function $C_{00}$ from Sec.~\ref{sec:c0} follows straightforwardly,
\begin{align}
  C_{00}(r) &:=
  \frac{r^2\rho^2-2r\rho+2-2\mathrm{e}^{-r\rho}}{4-4r\rho\mathrm{e}^{-r\rho}-2\mathrm{e}^{-2r\rho}}\cdot\mathrm{e}^{-r\rho}\, .
\end{align}

Figure~\ref{fig:1d_poisson} depicts the volume-volume correlation
function $C_{00}(r)$ of the one-dimensional Poisson point process;
both the analytic result and simulation data are shown.

As discussed in Sec.~\ref{sec:CmuPoisson} for the three-dimensional
Poisson point process, the Voronoi neighbors are correlated by
construction. Although very large Voronoi cells are rather unlikely,
their next-neighbor correlation leads to a large correlation length in
$C_{00}(r)$. In contrast to the three-dimensional case, the Voronoi
neighbors are uncorrelated if the distance of their centers vanishes
because in one dimension these Voronoi cells become independent. They
only depend on either the nearest neighbor on the left- or on the
right-hand side of $x_1=x_2$, which are independent of each other. For
large radii, the correlation vanishes exponentially, as expected,
because there is no long-range order in the Voronoi diagram.

\bibliographystyle{apsrev4-1.bst}
\bibliography{KlattTorquato2014}

\begin{thebibliography}{89}
\bibitem [{bibfootnote1()}]{bibfootnote1}%
  \BibitemOpen
  \bibinfo {note} {Note that there are many other more ordered packings with
  the same packing fraction.}\BibitemShut {Stop}%
\end{thebibliography}

\end{document}